\documentclass[12pt]{article}
\usepackage{epsf}

\renewcommand{\thefootnote}{\#\arabic{footnote}}
\begin{document}

\newcommand{\gtrsim}{ \mathop{}_{\textstyle \sim}^{\textstyle >} }
\newcommand{\lesssim}{ \mathop{}_{\textstyle \sim}^{\textstyle <} }

\renewcommand{\thefootnote}{\fnsymbol{footnote}}
\setcounter{footnote}{0}
\begin{titlepage}

\def\thefootnote{\fnsymbol{footnote}}

\begin{center}

\hfill TU-632\\
\hfill hep-ph/0110096\\
\hfill October, 2001\\

\vskip .5in

{\Large \bf
Effects of Cosmological Moduli Fields\\
on Cosmic Microwave Background
}

\vskip .45in

{\large
Takeo Moroi and Tomo Takahashi
}

\vskip .45in

{\em
Department of Physics, Tohoku University, Sendai 980-8578, Japan
}

\end{center}

\vskip .4in

\begin{abstract}

    We discuss effects of cosmological moduli fields on the cosmic
    microwave background (CMB).  If a modulus field $\phi$ once
    dominates the universe, the CMB we observe today is from the decay
    of $\phi$ and its anisotropy is affected by the primordial
    fluctuation in the amplitude of the modulus field.  Consequently,
    constraints on the inflaton potential from the CMB anisotropy can
    be relaxed.  In particular, the scale of the inflation may be
    significantly lowered.  In addition, with the cosmological moduli
    fields, {\sl correlated} mixture of adiabatic and isocurvature
    fluctuations may be generated, which results in enhanced CMB
    angular power spectrum at higher multipoles relative to that of
    lower ones.  Such an enhancement can be an evidence of the
    cosmological moduli fields, and may be observed in future
    satellite experiments.

\end{abstract}
\end{titlepage}

\renewcommand{\thepage}{\arabic{page}}
\setcounter{page}{1}
\renewcommand{\thefootnote}{\#\arabic{footnote}}
\setcounter{footnote}{0}

In superstring theory \cite{Polchinski}, it is well known that there
are various flat directions parameterized by scalar fields.
(Hereafter, we call these fields as ``moduli'' fields.)  Since their
potential is usually generated by effects of supersymmetry (SUSY)
breaking, their masses are expected to be of the order of the
gravitino mass.  Although masses of the moduli fields can be as light
as (or even lighter than) the electroweak scale, moduli fields do not
affect collider experiments since their interactions are suppressed by
inverse powers of the gravitational scale.

Cosmologically, however, they may cause serious problems
\cite{PRL131-59}.  Since the modulus field $\phi$ is a scalar field,
its primordial amplitude may be displaced from the minimum of the
potential, and this is naturally the case unless the minimum of the
potential is protected by some symmetry \cite{DinRanTho}.  If such a
displacement exists, then the modulus field starts to oscillate at
later stage of the universe and dominates the energy density of the
universe.  Since the interaction of the modulus field is expected to
be suppressed by inverse powers of the gravitational scale, its decay
width is at most
\begin{eqnarray}
    \Gamma_\phi \sim \frac{1}{4\pi} \frac{m_\phi^3}{M_*^2},
    \label{gamma_phi}
\end{eqnarray}
where $m_\phi$ is the mass of the modulus field and $M_*\simeq
2.4\times 10^{18}\ {\rm GeV}$ is the reduced Planck scale.  Using Eq.\ 
(\ref{gamma_phi}), $\phi$ decays before the present epoch if
$m_\phi\gtrsim O(100\ {\rm MeV})$.  In this case, the reheating
temperature is estimated as \cite{KolTur}
\begin{eqnarray}
    T_{\rm R} \simeq 1.2 g_*^{-1/4} \sqrt{M_* \Gamma_\phi}
    \sim 1.2\times 10^{-7}\ {\rm GeV} \times 
    \left( \frac{m_\phi}{100\ {\rm GeV}} \right)^{3/2},
    \label{T_R}
\end{eqnarray}
where $g_*$ is the effective number of the massless degrees of
freedom, and hence reheating temperature becomes lower than $\sim 1\
{\rm MeV}$ if $m_\phi\lesssim O(10\ {\rm TeV})$.  With such a low
reheating temperature, the great success of the standard big-bang
nucleosynthesis (BBN) is spoiled.  For lighter moduli fields
($m_\phi\lesssim O(100\ {\rm MeV})$), they survive until today and
overclose the universe.  Thus, the cosmological moduli fields cause
extremely serious problems in cosmology.

One solution to these difficulties is to push up the mass of the
moduli fields \cite{Heavyphi}.  In particular, in Ref.\
\cite{NPB570-455}, it was pointed out that the scenario with heavy
moduli can naturally fit into the framework of the anomaly-mediated
SUSY breaking \cite{AMSB}.  Indeed, according to Eq.\ (\ref{T_R}), the
reheating temperature can be higher than $\sim 1\ {\rm MeV}$ if
$m_\phi\gtrsim O(10\ {\rm TeV})$.  In this case, the BBN occurs after
the decay of the modulus field.  In this letter, we consider this
scenario and study its consequence in the cosmic microwave background
(CMB).

Although, in the case with the modulus field, the thermal history
after the BBN is mostly the same as the standard one, cosmology before
the modulus decay is completely different.  In particular, it should
be noted that the CMB we observe today is from the decay of the
modulus field while, in the conventional case, it is from the decay of
the inflaton.  Importantly, non-adiabatic fluctuation may be imprinted
in the initial amplitude of the modulus field during the inflation,
and we may observe a non-standard signal in the CMB angular power
spectrum.  Indeed, as we will see below, {\sl correlated} mixture of
adiabatic and isocurvature fluctuations may be induced, which results
in enhanced CMB angular power spectrum at higher multipoles relative
to that of lower ones.\footnote
{For other mechanism of generating correlated mixiture of adiabatic
and isocurvature fluctuations, see \cite{adi-iso}.}

To understand the effect of the cosmological modulus field, let us
follow the evolution of the modulus field from the epoch of the
inflation.  During the inflation, quantum fluctuation of the inflaton
field generates the source of the adiabatic perturbation.  Such an
effect is well parameterized by the gauge-invariant potential $\Psi$,
which is related to the perturbed line element in the Newtonian gauge:
\begin{eqnarray}
    ds^2 = - (1 + 2\Psi) dt^2 
    + a^2 (1 + 2\Phi) \delta_{ij} dx^i dx^j,
\end{eqnarray}
where $a$ is the scale factor.  (We use the notation of Ref.\ 
\cite{HuThesis}.)  Denoting $\Psi$ induced during the inflation as
$\Psi_{\rm i}$, we obtain \cite{PRD28-629}
\begin{eqnarray}
    \tilde{\Psi}_{\rm i} (k) =
    \left[ \frac{H_{\rm inf}^2}{2\pi |\dot{\chi}|}
    \right]_{k=aH_{\rm inf}},
\end{eqnarray}
where $k$ is the comoving momentum, $\chi$ is the inflaton field, the
``dot'' is the derivative with respect to time $t$, and $H_{\rm inf}$
is the expansion rate during the inflation. (Hereafter, the ``tilde''
is used for Fourier components of the correlation function with the
measure $\int d\ln k$.  For example, $\tilde{\Psi}(k)$ is defined as
$\langle\Psi(\vec{x})\Psi(\vec{y})\rangle=\int d\ln k
|\tilde{\Psi}(k)|^2 e^{i\vec{k}(\vec{x}-\vec{y})}$.)  If there is no
modulus field, this is the only source of the cosmic density
fluctuation.  However, if the modulus field exists, its amplitude may
also fluctuate, which can be a source of the density fluctuation.  If
the mass of the modulus is negligibly small relative to the expansion
rate during the inflation,\footnote
{If the mass of the modulus is comparable to or larger than $H_{\rm
inf}$, $\delta\tilde{\phi}$ becomes negligibly small and the
resultant CMB power spectrum is completely the same as the
conventional adiabatic case.  Therefore, we do not consider such a
case.}
we obtain
\begin{eqnarray}
    \delta \tilde{\phi}_{\rm i} (k) = \frac{H_{\rm inf}}{2\pi}.
    \label{dphi_i}
\end{eqnarray}
Hereafter, we assume that the initial amplitude of the modulus field
is large enough so that we can treat $\delta\tilde{\phi}$ as a
perturbation.  Importantly, two fluctuations $\tilde{\Psi}_{\rm i}$
and $\delta\tilde{\phi}_{\rm i}$ are uncorrelated and hence we can
study their effects separately.  Effects of $\tilde{\Psi}_{\rm i}$
have been intensively studied \cite{HuThesis}, and hereafter, we
concentrate on effects of $\delta\tilde{\phi}_{\rm i}$.

After inflation, inflaton field starts to oscillate and then decays.
Then, the universe is reheated and the radiation dominated universe is
realized.  (We call this epoch as ``RD1'' epoch.)  During this epoch,
the unperturbed modulus amplitude $\bar{\phi}$ and the perturbation in
$\phi$, denoted as $\delta\phi$, obey the following equations
of motion:
\begin{eqnarray}
    && \ddot{\bar{\phi}} + 3H \dot{\bar{\phi}}
    + V'(\bar{\phi}) = 0,
    \label{Eq-phi}
    \\ &&
    \ddot{\delta\tilde{\phi}} + 3H \dot{\delta\tilde{\phi}}
    + V''(\bar{\phi}) \delta\tilde{\phi} 
    + \frac{k^2}{a^2} \delta\tilde{\phi}
    = -2 \tilde{\Psi} V',
    \label{Eq-dphi}
\end{eqnarray}
where $H$ is the expansion rate of the universe, $V(\phi)$ is the
potential of $\phi$, and the ``prime'' is the derivative with respect
to $\phi$.  Since the modes we are interested in are at the
superhorizon scale during the RD1 epoch, we neglect the $k$-dependent
term in Eq.\ (\ref{Eq-dphi}) in the following discussion.  In
addition, we adopt the simplest potential for the modulus field:
\begin{eqnarray}
    V(\phi) = \frac{1}{2} m_{\phi}^2 \phi^2.
\end{eqnarray}
With this potential, notice that, in the long wavelength limit (i.e.,
$k\rightarrow 0$), $\bar{\phi}$ and $\delta\tilde{\phi}$ obey the same
equation if we neglect the gravitational potential $\Psi$.

It is instructive to discuss qualitative behaviors of $\bar{\phi}$ and
$\delta\tilde{\phi}$.  When $H\gg m_\phi$, $\bar{\phi}$ and
$\delta\tilde{\phi}$ both stay constant.  As the universe expands,
however, the expansion rate decreases and $H$ becomes comparable to
$m_\phi$ at some point.  Then, $\bar{\phi}$ and $\delta\tilde{\phi}$
both start to oscillate.  We assume that the reheating temperature
after the inflation is high enough so that the modulus field starts to
oscillate in the radiation dominated universe.  We also assume that
the initial amplitude of the modulus field $\bar{\phi}_{\rm i}$ is
smaller than $\sim M_*$; if this condition is satisfied, the energy
density of the modulus field is smaller than that of radiation when
$H\sim m_\phi$ is satisfied.

Once the modulus field starts to oscillate, equation of state for
$\phi$ becomes $\omega_\phi\rightarrow 0$ and $\phi$ behaves as a
non-relativistic matter.  Therefore, the situation is like the
conventional system with radiation and cold dark matter (CDM)
components with primordial isocurvature perturbation in the CDM
sector.  In particular, notice that the energy density of the modulus
is proportional to $a^{-3}$, and hence the energy density of the
modulus takes over that of radiation as the universe expands even
though initially $\rho_\gamma\gg\rho_\phi$, where $\rho_\gamma$ and
$\rho_\phi$ are energy densities of the radiation and modulus field,
respectively.  Therefore, at some point, the energy density of the
universe is dominated by that of the modulus field.  (We call this
epoch as ``modulus dominated'' or ``$\phi$D'' epoch.)

To study this system, it is convenient to define the entropy between
the modulus and the radiation:
\begin{eqnarray}
    S_{\phi\gamma} \equiv \frac{\delta\rho_\phi}{\rho_\phi}
    - \frac{3}{4} \frac{\delta\rho_\gamma}{\rho_\gamma},
\end{eqnarray}
where $\delta\rho_\gamma$ and $\delta\rho_\phi$ are fluctuations in
$\rho_\gamma$ and $\rho_\phi$, respectively.  Since $S_{\phi\gamma}$
is independent of time when $\omega_\phi$ becomes 0, we need to know
its initial value.  For this purpose, it is convenient to use the fact
that, for a given $\bar{\phi}(t)$ which is a solution to Eq.\
(\ref{Eq-phi}), $\delta\tilde{\phi}$ is given by
\begin{eqnarray}
    \delta\tilde{\phi} (k) = 
    \frac{\delta\tilde{\phi}_{\rm i} (k)}{\bar{\phi}_{\rm i}} 
    \bar{\phi}
    + \tilde{\Psi} (k) \dot{\bar{\phi}} t,
\end{eqnarray}
where we used an approximation that $\Psi$ is a constant, which is a
good approximation in the radiation dominated universe. In addition,
$\delta\tilde{\phi}_{\rm i}$ is the initial value of
$\delta\tilde{\phi}$ which is given in Eq.\ (\ref{dphi_i}).  We
identify the first and second terms as isocurvature and adiabatic
fluctuations in the modulus amplitude, respectively.  Indeed, with the
second term (with relevant perturbation in the radiation sector to
generate $\tilde{\Psi}_{\rm i}$), $S_{\phi\gamma}$ vanishes.  On the
contrary, with the first term, $S_{\phi\gamma}$ is given by
\begin{eqnarray}
    \tilde{S}_{\phi\gamma}(t) = \tilde{S}_{\rm i} (k) \equiv 
    \tilde{S}_{\phi\gamma}(t\rightarrow 0) =
    \frac{2 \delta\tilde{\phi}_{\rm i}(k)}{\bar{\phi}_{\rm i}}.
\end{eqnarray}
Notice that, if $\delta\tilde{\phi}_{\rm i}$ is independent of $k$,
$\tilde{S}_{\rm i}$ is also independent of $k$.

\begin{figure}[t]
    \centerline{\epsfxsize=0.75\textwidth\epsfbox{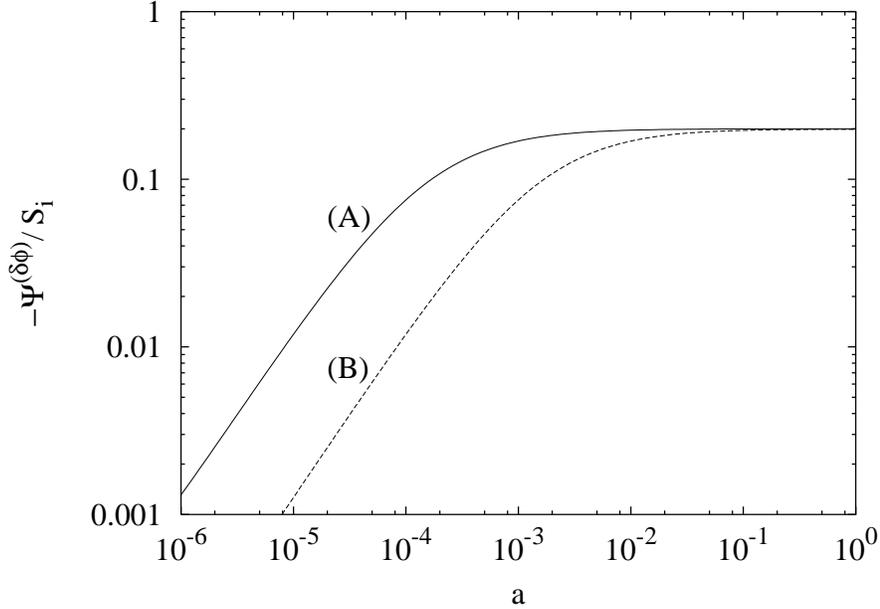}}
    \caption{Evolution of the gravitational potential at the 
    superhorizon scale for the case where $\delta\tilde{\phi}_{\rm
    i}\neq 0$ and $\tilde{\Psi}_{\rm i}=0$.  The horizontal axis is
    the scale factor $a$ whose normalization is arbitrary, and the
    vertical axis is $\tilde{\Psi}^{(\delta\phi)}$ normalized by
    $\tilde{S}_{\rm i}$.  The lines (A) and (B) correspond to cases
    with different initial modulus amplitude; $\bar{\phi}_{\rm i}^2$
    for the line (A) is 10 times larger than that of (B).}
    \label{fig:psi}
\end{figure}

We numerically followed the evolution of $\tilde{\Psi}^{(\delta\phi)}$
from the RD1 to the modulus dominated epoch.  In Fig.\ \ref{fig:psi},
we plot the evolution of the gravitational potential as a function of
the scale factor $a$.  Notice that, for the line (A) (line (B)), the
epochs with $a\lesssim 10^{-3}$ and $a\gtrsim 10^{-3}$ ($a\lesssim
10^{-2}$ and $a\gtrsim 10^{-2}$) correspond to the RD1 and modulus
dominated epochs, respectively.  It is interesting to compare cases
with different values of $\bar{\phi}_{\rm i}$.  As $\bar{\phi}_{\rm
i}$ decreases, the modulus dominated universe is realized at later
stage, as seen in Fig.\ \ref{fig:psi}.  However, the resultant value
of the ratio $\tilde{\Psi}^{(\delta\phi)}/\tilde{S}_{\rm i}$ is
independent of $\bar{\phi}_{\rm i}$ and is always $-0.2$.  We also
checked that the result is independent of $k$ as far as we consider
fluctuations at the superhorizon scale \cite{iso2adi}.

Behavior of $\tilde{\Psi}^{(\delta\phi)}$ given in Fig.\ \ref{fig:psi}
can be understood by denoting that the evolution of the density
fluctuations in this case is exactly like the conventional
isocurvature case.  When the universe is radiation dominated,
$\tilde{\Psi}^{(\delta\phi)}$ is proportional to $a$.  On the other
hand, once the universe is dominated by the non-relativistic matter
(i.e., the modulus field), $\Psi$ becomes the constant.  $\Psi$ at the
modulus dominated epoch is related to $\tilde{S}_{\rm i}$ as
\begin{eqnarray}
    \tilde{\Psi}^{(\delta\phi)}_{\rm \phi D}(k) 
    = - 0.2 \tilde{S}_{\rm i} (k),
    \label{Psi(phi-D)}
\end{eqnarray}
where the suffix $(\delta\phi)$ means that this quantity is induced by
$\delta\phi_{\rm i}$.  The relation (\ref{Psi(phi-D)}) holds until the
modulus field decays.

When the expansion rate of the universe becomes comparable to the
decay rate of $\phi$, the modulus field decays and energy density
stored in the modulus sector is transferred to that of radiation.
Then, a radiation dominated universe is again realized.  (We call this
epoch as ``RD2'' epoch.)  When $\phi$ decays, the density fluctuation
in the modulus sector is also transferred to the radiation.  This
results in a change in $\tilde{\Psi}$: with the decay, the equation of
state of the dominant component of the universe changes from $0$ to
$1/3$, and $\tilde{\Psi}$ varies by the factor of $10/9$:
\begin{eqnarray}
    \tilde{\Psi}^{(\delta\phi)}_{\rm RD2} (k) = 
    \frac{10}{9} \tilde{\Psi}^{(\delta\phi)}_{\rm \phi D}(k).
\end{eqnarray}
After the decay of the modulus field, we assume that scenario of the
standard cosmology (like the neutrino decoupling, standard BBN,
recombination, and so on) follows.

Now, we are at the position to discuss the CMB angular power spectrum
from the scenario with the modulus field.  First, we should emphasize
that there are two independent sources of the CMB anisotropy, i.e.,
$\tilde{\Psi}_{\rm i}$ and $\delta\tilde{\phi}_{\rm i}$.  Therefore,
the resultant CMB angular power spectrum is given in the following
form:
\begin{eqnarray}
    C_l = C_l^{\rm (adi)} + C_l^{(\delta\phi)}.
\end{eqnarray}
Here, $C_l^{\rm (adi)}$ is from the perturbation in the inflaton
field, which is of order $\tilde{\Psi}_{\rm i}^2$.  On the contrary,
$C_l^{(\delta\phi)}$ is from the primordial fluctuation of the modulus
amplitude, which is of order $\delta\tilde{\phi}_{\rm i}^2$.  (Notice
that there is no term which is of order $\tilde{\Psi}_{\rm
i}\delta\tilde{\phi}_{\rm i}$, since two fluctuations are
uncorrelated.)  $C_l^{\rm (adi)}$ can be calculated by following the
standard method.

In calculating $C_l^{(\delta\phi)}$, we must specify the origin of the
CDM and baryon.  Here, we assume that the CDM is generated by the
decay of the modulus field.  (For example, if the lightest
superparticle (LSP) is stable, decay of the modulus field may generate
right amount of the LSP for the CDM \cite{NPB570-455}.)  In this case,
after the decay of the modulus field, there is no entropy between the
CDM and radiation.

On the contrary, origin of the baryon asymmetry is more controversial.
Here, we consider two possibilities: (i) the baryon asymmetry is
(somehow) generated at the time of (or after) the decay of the modulus
field, or (ii) the Affleck-Dine (AD) mechanism \cite{NPB249-361}
generates the baryon number much before the decay of the modulus
field.

Let us first consider the case where the baryon number is generated by
the decay of the modulus field.  In this case, there is no entropy
between the baryon and the radiation, so the cosmic fluctuations are
exactly like the conventional adiabatic case once the modulus field
decays.  Thus, if we neglect the scale dependences of
$\tilde{\Psi}_{\rm i}$ and $\delta\tilde{\phi}_{\rm i}$,
$C_l^{(\delta\phi)}$ is proportional to $C_l^{\rm (adi)}$.  Therefore,
in this case, the CMB angular power spectrum is the same as the usual
adiabatic case if the normalization of the initial fluctuations are
properly chosen.  However, this fact has significant implications when
we construct a model of inflation. First of all, since
$\tilde{\Psi}^{(\delta\phi)}\sim O(H_{\rm inf}/\bar{\phi}_{\rm i})$,
we may generate large cosmic perturbation by lowering $\bar{\phi}_{\rm
i}$ even if $H_{\rm inf}$ is small.  Furthermore, usually, scale
dependence of $\delta\tilde{\phi}_{\rm i}$ is milder than that of
$\tilde{\Psi}_{\rm i}$.  Consequently, when $C_l^{\rm (adi)}\ll
C_l^{(\delta\phi)}$ is realized, the resultant CMB angular power
spectrum may be like that from the scale-invariant adiabatic
perturbation even if $\tilde{\Psi}_{\rm i}$ has a strong scale
dependence.  These facts relax constraints on the potential of the
inflaton field.  For example, this scenario provides an interesting
mechanism to lower the scale of inflation (i.e., $H_{\rm inf}$).

Now we consider the case where the baryon asymmetry is due to the AD
mechanism.  In this case, the amplitude of the AD field may fluctuate
and it can provide a new source of the cosmic perturbations.  In
general, perturbation in the AD field may be generated during the
inflation, which is $H_{\rm inf}/2\pi$ if the effective mass of the AD
field during inflation is much smaller than $H_{\rm inf}$.  If such a
fluctuation exists, it becomes a source of an {\sl uncorrelated}
isocurvature perturbation in the baryonic sector.  To make our point
clearer, we assume that the initial value of the fluctuation in the AD
field is negligibly small. This may happen when, for example, the
effective mass of the AD field is comparable to $H_{\rm inf}$
during the inflation.  It should be also noted that, if the initial
amplitude of the AD field is much larger than $\bar{\phi}_{\rm i}$,
effects of the fluctuations in the AD field becomes also negligible.
(Effects of the fluctuations in the AD field will be discussed
elsewhere \cite{MTinProgress}.)

Even when there is no primordial fluctuation in the AD field, however,
non-vanishing isocurvature fluctuation is eventually generated in the
baryonic sector if $\delta\tilde{\phi}_{\rm i}\neq 0$.  If we
calculate the entropy between the baryon and $\phi$ before the decay
of the modulus field, we obtain
\begin{eqnarray}
    \tilde{S}_{b\phi} \equiv
    \frac{\delta\tilde{\rho}_b}{\rho_b} - 
    \frac{\delta\tilde{\rho}_\phi}{\rho_\phi} =
    - \tilde{S}_{\rm i},
\end{eqnarray}
where we have used the fact that $\delta\tilde{\rho}_b\rightarrow 0$
as $a\rightarrow 0$.  This entropy is conserved until the modulus
field decays, and it becomes the entropy between the photon and baryon
after the decay of the modulus field.  Therefore, in calculating
$C_l^{(\delta\phi)}$ which is generated by the primordial fluctuation
in the modulus amplitude, initial condition for the baryonic density
fluctuation is different from the conventional adiabatic case, and is
given by, in deep radiation dominated epoch,
\begin{eqnarray}
    \left. \frac{\delta\tilde{\rho}_b}{\rho_b} \right|_{\rm RD2} = 
    -\tilde{S}_{\rm i} 
    + \frac{3}{4}
    \left. \frac{\delta\tilde{\rho}_\gamma}{\rho_\gamma} 
    \right|_{\rm RD2} =
    4.5 \tilde{\Psi}^{(\delta\phi)}_{\rm RD2} + \frac{3}{4}
    \left. \frac{\delta\tilde{\rho}_\gamma}{\rho_\gamma} 
    \right|_{\rm RD2}.
    \label{db_i}
\end{eqnarray}
Notice that the first term in the right-handed side of the above
equation does not exist in the usual adiabatic ones.  Initial
conditions for other perturbations are the same as the usual adiabatic
case.  The first term in Eq.\ (\ref{db_i}) gives rise to the
non-vanishing entropy in the baryonic sector and hence we call it as
the ``isocurvature'' term.  Here, it should be emphasized that such an
isocurvature fluctuation is {\sl correlated} with the contribution
from $\tilde{\Psi}^{(\delta\phi)}_{\rm RD2}$ which gives rise to the
effect like the conventional adiabatic fluctuation.  Thus, the effect
of this ``isocurvature'' fluctuation is completely different from the
conventional {\sl uncorrelated} isocurvature fluctuation and, as we
will see below, its effect may be observable in future satellite
experiments.

\begin{figure}[t]
    \centerline{\epsfxsize=0.75\textwidth\epsfbox{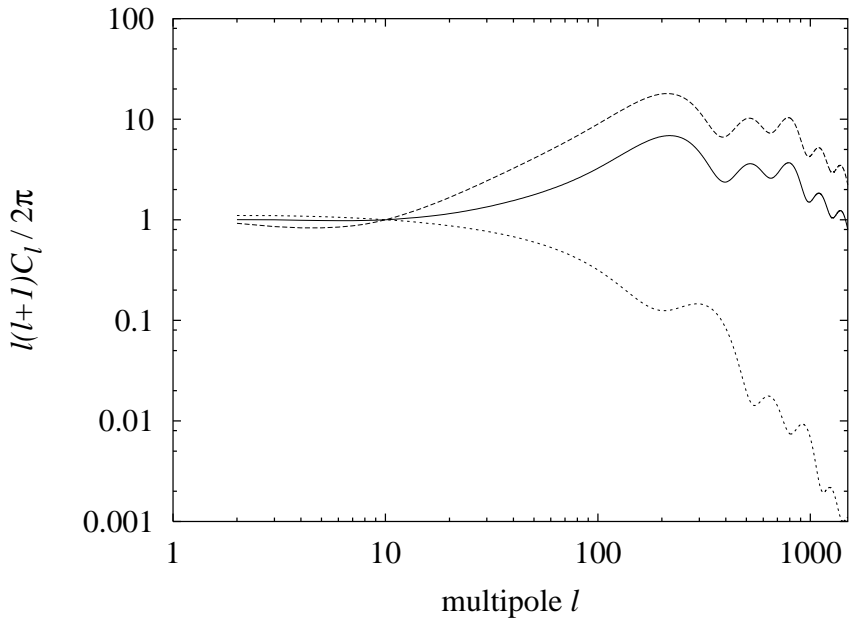}}
    \caption{The CMB angular power spectrum $C_l^{(\delta\phi)}$ 
    for the case with the correlated isocurvature perturbation in the
    baryonic sector (dashed line), as well as $C_l$ for purely
    adiabatic (solid line) and purely baryonic isocurvature (dotted
    line) cases.  For the cosmological parameters, we use $h=0.65$,
    $\Omega_{\rm b}h^2=0.019$, $\Omega_{\rm m}=0.4$ \cite{aph0007187},
    and the flat universe is assumed.  (Here, $h$ is the present
    expansion rate of the universe in units of 100 km/sec/Mpc, and
    $\Omega_{\rm b}$ and $\Omega_{\rm m}$ are present density
    parameters for baryon and non-relativistic matter, respectively.)
    We used the normalization $[l(l+1)C_{l}/2\pi]_{l=10}=1$.}
    \label{fig:comparison}
\end{figure}

In Fig.\ \ref{fig:comparison}, we show $C_l^{(\delta\phi)}$ for the
case with the {\sl correlated} isocurvature perturbation in the
baryonic sector.  For comparison, we also plot the angular power
spectra with purely adiabatic perturbation and with purely baryonic
isocurvature perturbation.  As we can see, $C_l^{(\delta\phi)}$ is
completely different from other two cases; for the case with
correlated isocurvature perturbation, the CMB angular power spectrum
at higher multipoles is enhanced relative to that for the lower ones
compared to other cases \cite{PRD62-043504}.  Thus, if the total
angular power spectrum has some contamination of $C_l^{(\delta\phi)}$,
$C_l$ at higher multipoles is enhanced relative to that at lower ones.
Notice that such a signal is distinguishable from that with {\rm
uncorrelated} isocurvature perturbation, and that it may provide an
interesting evidence of the cosmological moduli fields.

Now, we show the total CMB angular power spectrum with such a
correlated isocurvature fluctuation in the baryonic density
fluctuation.  Since there are two sources of the cosmic perturbations,
$\Psi_{\rm i}$ and $S_{\rm i}$ (or equivalently, $\delta\phi_{\rm
i}$), we define\footnote
{For simplicity, we neglect the scale dependence of $R$.}
\begin{eqnarray}
    R \equiv \frac{S_{\rm i}}{\Psi^{\rm (adi)}_{\rm RD2}},
\end{eqnarray}
where $\Psi^{\rm (adi)}_{\rm RD2}$ is the gravitational potential in
the adiabatic mode during the RD2 epoch, which is related to
$\Psi_{\rm i}$ as
\begin{eqnarray}
      \Psi^{\rm (adi)}_{\rm RD2} = \frac{4}{9} \Psi_{\rm i}.
\end{eqnarray}
For the case of $\bar{\phi}_{\rm i}\sim M_*$, $R$ is expected to be
$O(1)$ when $\tilde{\Psi}_{\rm i}\sim O(H_{\rm inf}/M_*)$, which is
the case in, for example, the chaotic inflation models.  However, in
general, $R$ can be much larger or smaller than $1$ depending on the
model of the inflation as well as on $\bar{\phi}_{\rm i}$, and hence
we treat $R$ as a free parameter.

\begin{figure}
    \centerline{\epsfxsize=0.75\textwidth\epsfbox{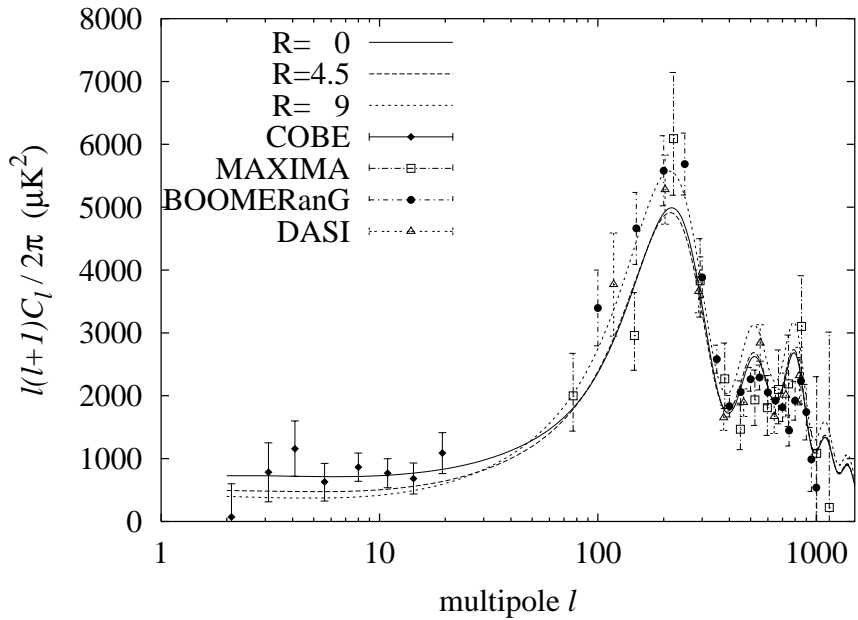}}
    \caption{The CMB angular power spectrum $C_l$ for 
    $R=0$ (solid), $R=4.5$ (dashed), and $R=9$ (dotted).  The overall
    normalization of $C_l$ is determined such that the $\chi^2$
    variable is minimized.  Scale-invariance is assumed both for
    $\Psi_{\rm i}$ and $S_{\rm i}$.  The values of the cosmological
    parameters are the same as those used in Fig.\ 
    \ref{fig:comparison}.}
    \label{fig:C_l}
\end{figure}

In Fig.\ \ref{fig:C_l}, we plot the total CMB angular power spectrum
for several values of $R$.  Here (and in the following analysis), the
overall normalization of $C_l$ is determined such that the
goodness-of-fit parameter $\chi^2=-2\ln L$, where $L$ is the
likelihood function, is minimized.  In our analysis, following Ref.\ 
\cite{APJ533-19}, the offset lognormal approximation is used.  In
addition, we take account of the data from COBE \cite{cobe}, BOOMERanG
\cite{boomerang}, MAXIMA \cite{maxima}, and DASI \cite{dasi}.  For our
numerical analysis, we use {\tt RADPACK} package \cite{radpack} to
calculate $\chi^2$, which is based on 24, 19, 13, and 9 band powers
from COBE, BOOMERanG, MAXIMA, and DASI, respectively.\footnote
{For the BOOMERanG and MAXIMA data, the cross-correlations between the
band powers are not included in our analysis since they are not
available.}
As one can see, correlated isocurvature perturbation in the baryonic
sector may result in an enhancement of $C_l$ at higher multipoles
relative to that at lower ones.  Therefore, future satellite experiment
will give us interesting tests of the scenario with the cosmological
moduli fields.

\begin{figure}
    \centerline{\epsfxsize=0.75\textwidth\epsfbox{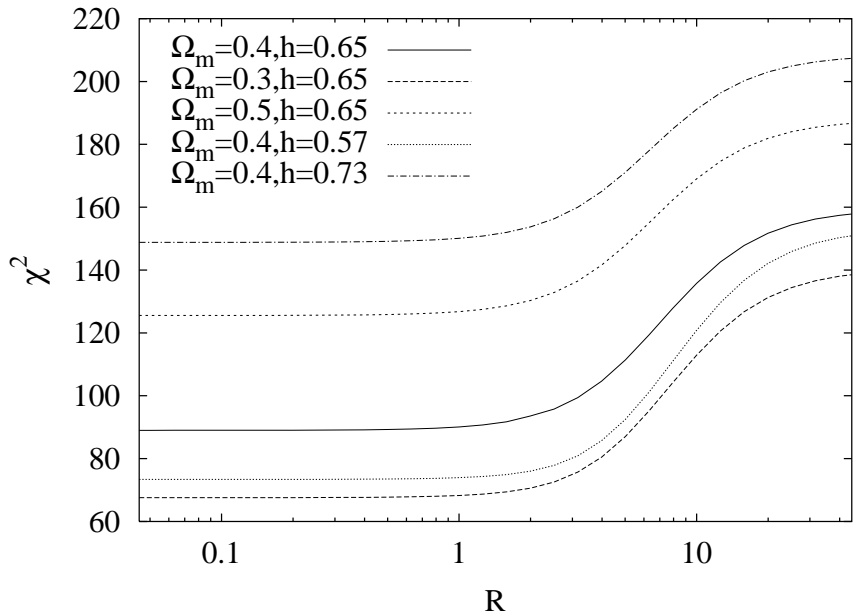}}
    \caption{The $\chi^2$ variable as a function of $R$.  The overall
    normalization of $C_l$ is chosen so that the $\chi^2$ variable is
    minimized.  We take $\Omega_{\rm b}h^2=0.019$, and several values
    of $h$ and $\Omega_{\rm m}$ \cite{aph0007187}.}
    \label{fig:chi^2}
\end{figure}

In fact, even the existing data for the CMB power spectrum provides a
constraint on the parameter $R$.  Since $C_l$ at higher multipoles
becomes larger as $R$ increases, too large $R$ is excluded.  In Fig.\ 
\ref{fig:chi^2}, we plot the $\chi^2$ variable as a function of $R$,
where we take $\Omega_{\rm b}h^2=0.019$, and several values of $h$ and
$\Omega_{\rm m}$.  As one can see, the value of $\chi^2$ is sensitive
to the cosmological parameters since the heights and locations of the
acoustic peaks depend on them.  For example, taking $(h, \Omega_{\rm
b}h^2, \Omega_{\rm m})=(0.65, 0.019, 0.3)$, which results in the most
conservative upper bound on $R$ among the data sets used in Fig.\ 
\ref{fig:chi^2}, and requiring $\chi^2\leq 84$, which gives 95 \% C.L.
allowed region for the $\chi^2$ statistics with 64 degrees of freedom,
we obtain the constraint $R\leq 4.5$.  Thus, we can exclude the case
where the CMB anisotropy is purely from the primordial fluctuation in
the modulus amplitude (i.e., the case of $R\rightarrow\infty$).

In summary, we have studied the effects of the cosmological moduli
fields on the CMB anisotropy.  Importantly, if a cosmological modulus
field exists, it may significantly affect the CMB.  In particular, in
the scenario with the cosmological moduli fields, {\sl correlated}
isocurvature fluctuation may exist in the baryonic sector which
results in enhanced CMB angular power spectrum at higher multipoles
relative to that at lower ones.  Such an effect can be a striking
signal from the cosmological moduli fields, and it may be observed in
future satellite experiments. In addition, even in the case where
there is no isocurvature perturbation, the cosmological modulus field
may have important implication to the model-building of the inflation;
it may relax constraints on the inflaton potential.  In particular,
the cosmological modulus field provides an interesting possibility to
lower the scale of the inflation adopting smaller value of the initial
amplitude of the modulus field.  In fact, similar mechanism of
changing the constraints on the inflaton potential may work with the
AD field; in this case the baryon asymmetry is from the decay of the
AD field and the correlated isocurvature perturbation in the baryonic
sector is absent.  More detailed discussion in this case will be given
elsewhere \cite{MTinProgress}.

{\sl Note added:} While preparing the manuscript, we found a paper by
D.H. Lyth and D. Wands \cite{hph0110002}, which has some overlap with
our analyses.

{\sl Acknowledgment:}  This work is supported by the Grant-in-Aid for
Scientific Research from the Ministry of Education, Science, Sports,
and Culture of Japan, No.\ 12047201 and No.\ 13740138.

\end{document}